\documentclass[3p,times,procedia]{elsarticle}

\usepackage{amssymb}

\usepackage[figuresright]{rotating}

\begin{document}

\begin{frontmatter}




\title{Improved jet clustering algorithm with vertex information for multi-bottom final states}


\author{Taikan Suehara}
\author{Tomohiko Tanabe}
\author{Satoru Yamashita}

\address{International Center for Elementary Particle Physics, The University of Tokyo, 7-3-1 Hongo, Bunkyo-ku, Tokyo, 113-0033 Japan}

\begin{abstract}
In collider physics at the TeV scale,
there are many important processes which involve six or more jets.
The sensitivity of the physics analysis depends critically
on the performance of the jet clustering algorithm.
We present a full detector simulation study for the ILC
of our new algorithm which makes use of secondary vertices
which improves the reconstruction of $b$ jets.
This algorithm will have many useful applications,
such as in measurements involving
a light Higgs which decays predominantly into two $b$ quarks.
We focus on the measurement of the Higgs self-coupling,
which has so far proven to be challenging but is
one of the most important measurements at the ILC.
\end{abstract}

\begin{keyword}
jet clustering \sep vertex finder \sep Higgs self coupling
\PACS 07.05.Kf \sep 29.85.Fj

\end{keyword}

\end{frontmatter}


\section{Introduction}
\label{sec:introduction}

Jet clustering is an essential technique in high energy physics experiments
in which the multitude of produced particles are combined into \emph{jets}
which represents an attempt to reconstruct the originating
quarks and gluons in the final state.
The development of jet clustering algorithm has a long history
ever since QCD jets have been produced in particle collisions;
we name a few examples in lepton colliders
such as the Jade algorithm~\cite{Bartel:1986ua}, the Durham algorithm~\cite{Catani:1991hj},
and the Cambridge algorithm~\cite{Dokshitzer:1997in}.
These algorithms have dealt with the challenging question of
how to deal with gluon emissions of various energies.
With the advent of future lepton colliders at the TeV scale,
the number of quarks in the final state increases roughly with the collision energy,
which makes even more challenging to correctly group the resulting hadrons
into their originating partons.
There is further complication arising from
the imbalance of the parton energies due to
the difference in their origin, such as whether they
directly come from $e^+e^-$ collisions or from $W$ or $Z$ boson decays,
and also because of initial state radiation which adds a boost to the system.

We will focus on the physics application of jet clustering at a future lepton collider,
such as the International Linear Collider (ILC),
although applications to hadron colliders should be possible with minor adjustments.

The ability to group the particles according to their originating parton
is particularly important in the analysis of physics processes
involving multi-jet final states, such as
the measurement of the Higgs self-coupling,
which uses the $e^+e^-\to ZHH$ channel for $\sqrt{s}=500$~GeV,
or the top Yukawa coupling, which uses the
$e^+e^-\to t\overline{t}H \to bW^+\overline{b}W^-H$ channel.
Depending on the decay modes of the $W$, $Z$, and the Higgs,
the number of jets in the final state can be as high as
6 for $ZHH$ and 8 for $t\overline{t}H$.
This is relevant especially in
the case of a light Higgs particle
as motivated by electroweak precision measurements,
whose branching ratio of $H\to b\overline{b}$
is 68\% for a Higgs mass of 120~GeV.
These channels have major background processes with similar number of jets;
particularly important is
$e^+e^-\to t\overline{t}$ whose has a large cross section.
Many such processes can be greatly reduced
if the flavor of the originating quark can be identified;
by requiring the correct number of reconstructed
jets originating from $b$ quarks (``$b$ jets''),
the $t\overline{t}$ background could be eliminated.
In reality, flavor identification itself is a challenging task,
which results in a leakage of $t\overline{t}$ background
even with an efficient flavor identification algorithm,
which results in significant background due to the sheer size of the cross section.
Other backgrounds include those in which
the Higgs decay $H\to b\overline{b}$
is replaced by the $Z$ decay $Z\to b\overline{b}$,
which becomes an irreducible background.

Flavor identification can be accomplished by looking for signs of secondary decays
of $b$ hadrons whose proper lifetime is typically 400-500~$\mu $m$/c$.
This results in, for example,
a heavy tail in the impact parameter distributions of charged tracks,
secondary vertices which are displaced from the primary vertex,
increased transverse momentum relative to the jet direction
due to the heavy $b$ hadron, as well as 
the presence of leptons due to semileptonic decays of the $W$ boson.
Such signatures are typically combined using
a multivariate analysis technique \cite{Bailey:2009ui}
into a single variable which can be used to
discriminate $b$ jets from jets originating from lighter quarks.
Similar techniques can be applied to identify $c$ jets.

Traditionally, the jet clustering procedure is performed first,
after which the flavor identification algorithm is applied
to each of the resulting jets.
The search for secondary vertices is restricted to
combination of particles within the jet,
which reduces the computing cost arising from combinatorial effects.
This method has the consequence that mistakes in jet clustering,
such as particles originating from the same vertex
being associated into separate jets (vertex splitting),
or the inclusion of multiple vertices of $b$ origin
into a single jet (vertex merging),
cannot be fixed at a later stage.
As computing resources grow inexpensive,
performing the vertex finding procedure
using all particles in the event
can be performed in a reasonable amount of computing time.
Our methods exploit this fact and use it to
improve the jet clustering procedure.
In this study, we show that, in multi-jet environment,
the accuracy of jet clustering is significantly improved by this method.

\section{Framework}
\label{sec:framework}
The software framework used in this study is based on LCIO~\cite{Gaede:2003ip}.
The detector simulation is performed by Mokka, a Geant4 based program.
Collisions of electron and positron beams are simulated with
the International Large Detector (ILD) Concept~\cite{:2010zzd}
at a center-of-mass energy of 500~GeV.
Initial state radiation and beamstrahlung effects are included.
The event reconstruction is done using the Marlin framework,
which consists of a series of modules which perform
hit digitization and smearing, track finding,
and particle flow analysis (PFA) using the Pandora algorithm~\cite{Thomson:2009rp}.
Neutral clusters are identified as a result of PFA.
The jet clustering is performed using various algorithms,
including the one developed for this study.
Flavor tagging is performed by using the LCFIVertex algorithm~\cite{Bailey:2009ui}.

For performance studies, we use a sample of $e^+e^-\to ZHH \to qqbbbb$ events,
which we consider as signal,
and a sample of $e^+e^-\to t\overline{t} \to bbcssc$ events
as a representative background
which illustrates the power of the jet clustering algorithms;
this study is by no means a comprehensive physics analysis.
Events are generated assuming a Higgs mass of 120~GeV and a top quark mass of 175~GeV.
We generate about 10000 events with the ILD full simulation framework for each process.


\section{Methods}

\subsection{Basics}

There are many jet clustering algorithms used in collider experiments.
Many of these algorithms begin by treating
every particle (track or calorimeter cluster)
as a jet on its own right.
Each jet is combined with one another, based on the criteria defined by the algorithm,
until either a certain threshold is reached
or the desired number of jets is obtained.
We focus on the case where the number of jets is reduced by one in each step,
which is the case for the Durham algorithm~\cite{Catani:1991hj} described below.
At each step of the algorithm,
a distance measure $Y(i,j)$, for the $i$-th and $j$-th jets,
is computed for every pair of jets.
The pair which has the smallest $Y$ value is combined into a single jet.
The Durham algorithm uses the distance measure defined as
\begin{equation}
Y(i,j) = \frac{2\,\mathrm{min}\left(E_i, E_j\right)^2(1-\cos\theta_{ij})}{Q^2},
\end{equation}
where $E_i$ and $E_j$ stand for the jet energies,
and $\theta_{ij}$ is the angle between the two jets.
The specific energy, which is constant for all events, is given as $Q^2$,
which is typically the center-of-mass energy.
Since the Durham algorithm gives a good performance
in a wide range of event topologies,
we use it to compare with our new jet clustering algorithm.

The algorithm aims to separate the particles which originate from different heavy hadrons
as well as to combine the particles which originate from the same heavy hadron.
For this purpose,
we incorporate the information from secondary vertices as well as particle identification,
in contrast to the existing methods which work primarily with the 4-momenta of the particles.
The crucial step is in identifying the signatures of heavy hadrons
before performing the jet clustering.
Below, we give a detailed description of our method, which consists of the following steps:
vertex finding, vertex selection, lepton finder, vertex combination and jet clustering.

\subsection{Vertex finder and selection}

In the proposed method, the vertex finder drives the performance of the jet clustering,
since any fake vertex degrades the performance
and is no better than the existing jet clustering methods.
At the same time, one needs a sufficiently high vertex reconstruction efficiency
to make an impact.
Thus we require a vertex finder which is optimized toward
high purity and with competitive reconstruction efficiency.

There are many vertex finders which are used to identify
secondary vertices in heavy-flavor jets.
Since they have been used after the jet clustering step,
they often have optimizations which take into account the jet direction.
This is the case, for instance, for the
topological vertex finder, the ZVTOP algorithm~\cite{Jackson:1996sy},
as implemented in the ILD full simulation framework.
Instead of adapting it to our purpose,
we have developed our own vertex finders based on existing techniques
which have been optimized to match our goals.

We adopt two methods for vertex finding,
one for the primary vertex and the other for the secondary vertices.
For the primary vertex finder,
we use the \emph{tear-down} type method,
while for the secondary vertices
we use the \emph{build-up} type method.
Both vertex finders are based on the simple vertex fitter,
which is implemented as follows.
Using Minuit2~\cite{James:1975dr},
we fit for the point in three-dimensional space
which minimizes the $\chi^2$ value
computed by the distance between each track and the point,
divided by the error given by the track covariance matrix,
summed for all tracks which are being considered for the vertex.
The initial condition for the fit is given by a simple geometrical calculation of
the closest point to all tracks without taking into account the errors.

The primary vertex finder begins by taking all tracks in the event.
They become the list of primary track candidates.
The $\chi^2$ value is then computed for every track.
The track which has the largest contribution to the $\chi^2$ value
is dropped from the list of primary track candidates.
The vertex is then refitted with the new list of primary tracks.
This procedure is repeated until each track has
a $\chi^2$ contribution of less than 25.

The secondary vertex finder begins by considering all tracks
which are not associated with the primary vertex (non-primary tracks).
Here, the \emph{build-up} strategy is applied,
so that we first form pairs using the non-primary tracks.
A tight quality selection is applied to these initial pairs
by requiring the $\chi^2$ value of less than 9 and
applying selections on the vertex mass and the combined momentum direction.
The refined pairs become the initial vertices
and are then considered for merging with other tracks.
We loop over the non-primary tracks to test against each vertex.
For each new trial track,
the $\chi^2$ value is recomputed including the new track.
If the resulting vertex passes the same quality selection described above,
the new vertex is retained.
This process is repeated until no other tracks can be attached.
At the end, checks are performed to eliminate duplicate vertices and multiple uses of tracks.
Priorities are given based on the
number of tracks in the vertex and the $\chi^2$ probability of the vertex.

The resulting secondary vertices are passed through another round of quality selection
which aims to reduce $V^0$ and fake vertices.
Vertices which have a mass consistent with that of $K^0_S$ are rejected.
Vertices which are too far ($>$ 30 mm) or
too near ($<$ 0.3 mm) from the primary vertex are also eliminated.

The performance of our secondary vertex finder is compared with that of ZVTOP
based on the origin of tracks using the information from the event generator.
Here, we categorize all tracks using the generator information
into the following classes:
(1) primary tracks, (2) $b$ track, 
(3) $c$ track, and
(4) other tracks,
which include decays from $\tau$, $K^0_S$, and conversions.
We count the number of tracks used by the secondary vertex finder.
In the result summary shown in Table~\ref{tbl:vertex_performance},
we define \emph{good} vertices as those whose tracks
come from the same heavy hadron, but not requiring 
to come directly from the same parent.
For the result of ZVTOP, we apply the Durham jet clustering constrained to 6 jets.
The result for our original vertex finder does not use jet clustering.
This comparison shows that our original vertex finder is
comparable in efficiency (slightly less efficient) to ZVTOP,
while the rejection rate of primary tracks and other tracks
is far superior for the tested samples of 6 jet events.

\begin{table}
\begin{minipage}[t]{.48\textwidth}
\begin{center}
\small{
\begin{tabular}{|l|r|r|r|r|r|}\hline
 & & \multicolumn{2}{c|}{ZVTOP} & \multicolumn{2}{c|}{Original} \\ \cline{3-6}
Category & All & All & Good & All & Good \\ \hline\hline
Primary & 10731 &  160 &    - &   54 &    - \\ \hline
$b$     &  2037 & 1399 & 1344 & 1309 & 1303 \\ \hline
$c$     &  2433 & 1653 & 1618 & 1571 & 1562 \\ \hline
Others  &   587 &  159 &   45 &   46 &   18 \\ \hline
\end{tabular}
}

(a) $qqbbbb$ sample
\end{center}
\end{minipage}
\begin{minipage}[t]{.48\textwidth}
\begin{center}
\small{
\begin{tabular}{|l|r|r|r|r|r|}\hline
 & & \multicolumn{2}{c|}{ZVTOP} & \multicolumn{2}{c|}{Original} \\ \cline{3-6}
Category & All & All & Good & All & Good \\ \hline\hline
Primary &  6980 &   76 &    - &   14 &    - \\ \hline
$b$     &   893 &  612 &  593 &  579 &  573 \\ \hline
$c$     &  1627 & 1086 & 1052 & 1045 & 1035 \\ \hline
Others  &   430 &  119 &   28 &   53 &   19 \\ \hline
\end{tabular}
}

(b) $bbcssc$ sample
\end{center}
\end{minipage}
\caption{Comparison of the performance of ZVTOP and our original vertex finder.
The numbers in each method show the number of tracks associated with
the reconstructed vertices.
If all the tracks in a vertex come from the same
heavy hadron, the vertex is counted as \emph{good}.}
\label{tbl:vertex_performance}
\end{table}

\subsection{Lepton finder}

Isolated leptons within a jet can be a sign of
semileptonic decays of heavy flavor hadrons.
Here, we focus on muons instead of electrons,
since electron identification suffers from
the incorrect matching of calorimeter clusters with the track.
We use a simple muon selection criteria
by requiring an energy deposit of greater than 50 MeV in the muon chamber,
while limiting the energy deposits inside the
electromagnetic and hadron calorimeters.
To further increase the purity of the muon selection,
we require the impact parameter of the track
in either direction ($d_0$ or $z_0$) to be displaced from
the primary vertex by larger than 5 $\sigma$.
These muons are treated in equal footing as secondary vertices
in the procedure below.

\subsection{Vertex and lepton combination}

A striking feature of heavy flavor hadrons is the cascade of multiple decays.
The purpose of this step is to combine
the secondary vertices and the leptons from the semileptonic decays
in a way that is consistent with the cascade decay.
The combination is done using the opening angles between the vertices and/or leptons.
For the vertex, the direction of the vertex position from the primary vertex is used,
while for the leptons the momentum direction is used.
A pair of two vertices are combined if the opening angle between the two vertices is
less than 0.2~rad.
For a pair of two leptons or a lepton and a vertex,
the opening angle threshold is 0.3~rad,
considering the fact that leptons tend to have a larger deviation in angle
with respect to the jet direction.

\subsection{Jet clustering}

The jet clustering is the last step of our method.
First, the vertices and leptons are treated as jet cores.
If the number of jet cores is larger than the required number of jets,
the nearest jet cores are combined until the required number is reached.
The resulting jet cores are kept separate in the procedure below.

Second, the remaining tracks and neutral clusters,
including those that come from the primary,
are combined to one of the jet cores.
We perform this in two steps, first with a cone jet clustering algorithm
and then with the traditional Durham-like clustering algorithm.
By looking at the opening angle between the momentum direction of the particle
and that of the jet cores,
those which fall within 0.2 radian of the jet core are merged with that jet core.
If there are multiple possible jet cores to combine,
the one with the closest jet core is used.
The remaining particles (tracks or clusters) are
combined to the jet cores based on the Durham $Y_{ij}$ distance measure.
In this step, we prevent the jet cores from merging with each other.


\section{Results}

All results in this section use the jet clustering with six jets,
    both for Durham and our jet clustering algorithm.

\subsection{Number of jets with $b$ hadrons in $ZHH \rightarrow bbbbbb$ events}

Here, we use the six $b$ sample, extracted as a subset of $ZHH$ events.
In this sample, every reconstructed jet must include one and only one $b$ hadron
if the jet clustering is done perfectly.
Therefore, counting the number of jets which include at least one $b$ hadron
is a good performance test for this process.

The $b$ hadrons are identified using MC generator information.
Each $b$ hadron is associated to a jet which has the largest number
of tracks from the $b$ hadron.
After associating all the $b$ hadrons,
we count the number of jets containing the $b$ hadrons.

\begin{figure}
\begin{center}
\includegraphics[width=.5\textwidth]{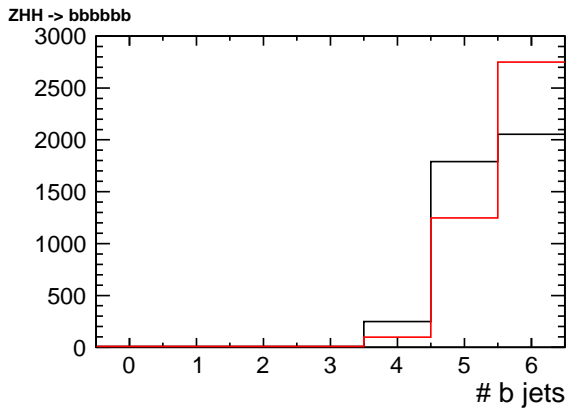}
\caption{Number of reconstructed jets including $b$ hadron
  with $ZHH \to bbbbbb$ events in each method.
The red line shows the result of our original algorithm
and the black line shows the result of Durham algorithm.}
\label{fig:countbhadron}
\end{center}
\end{figure}

Figure~\ref{fig:countbhadron} shows the result with both Durham and our original method.
The fraction of events which give all jets associating to $b$ hadrons
is increased from 52\% to 66\% by using our method instead of the Durham method.

\subsection{Number of $b$-hadron tracks in each jet}

In all of the following studies,
we focus on the separation of the $ZHH \to qqbbbb$ signal
from the $t\overline{t} \to bbcssc$ background.
In this study, we count the tracks from $b$ hadrons in each reconstructed jet.
Again, the $b$ hadrons and their daughter tracks are identified using
MC generator information.

Since the number of $b$ jets is usually 4 in the $qqbbbb$ process
and 2 in the $bbcssc$ process,
the number of $b$-hadron tracks can be a good separation criteria.
After ordering jets with descending order of number of $b$-hadron tracks,
we examine the numbers of $b$-hadron tracks in third and fourth jets.
They are expected to be zero in $bbcssc$ and non-zero in $qqbbbb$
if the jet clustering is done perfectly.

\begin{figure}
        \begin{center}
        \begin{minipage}[t]{.48\textwidth}
        \begin{center}
    \includegraphics[width=.8\textwidth]{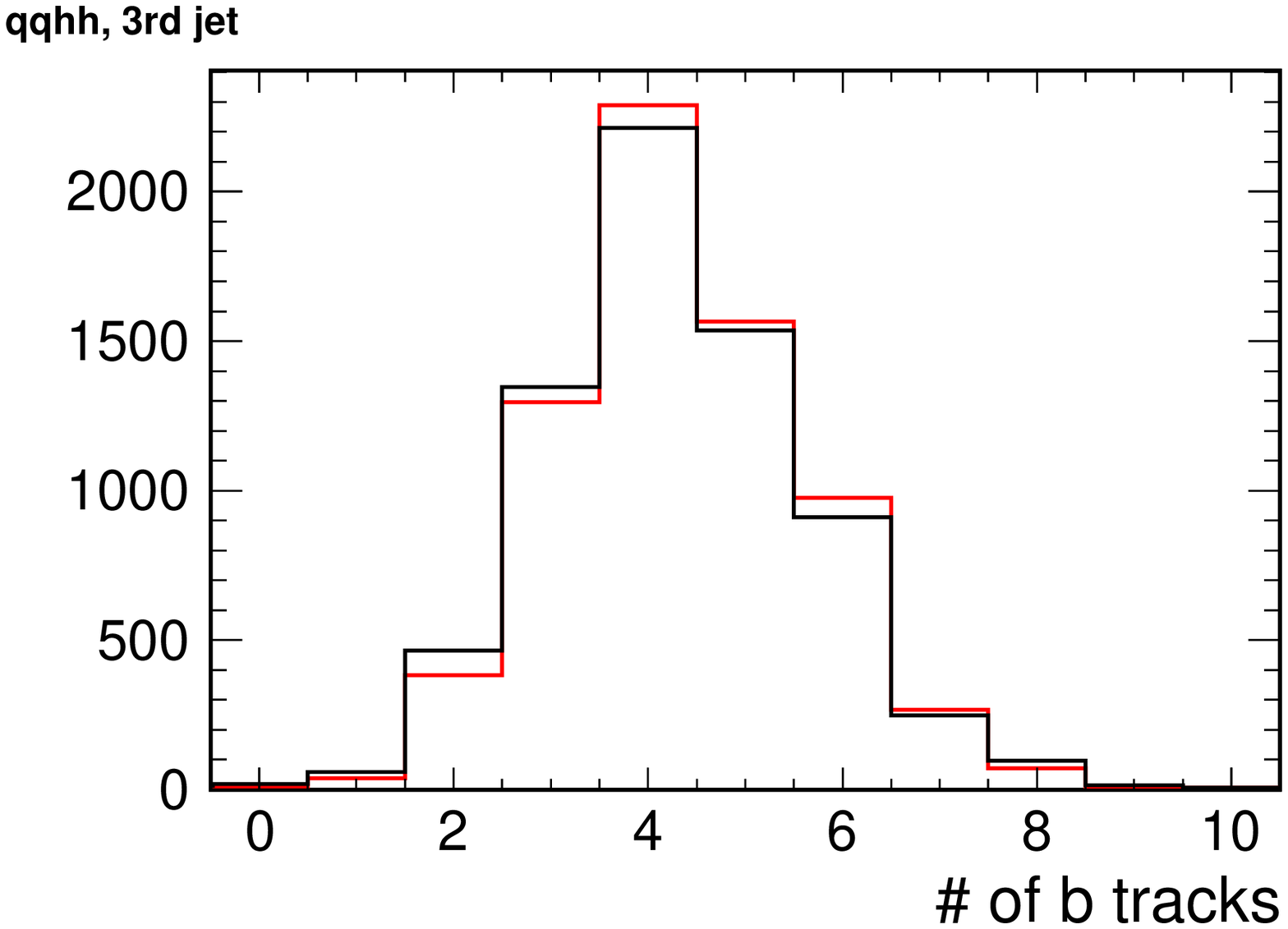}
        \end{center}
        \end{minipage}
        \begin{minipage}[t]{.48\textwidth}
        \begin{center}
    \includegraphics[width=.8\textwidth]{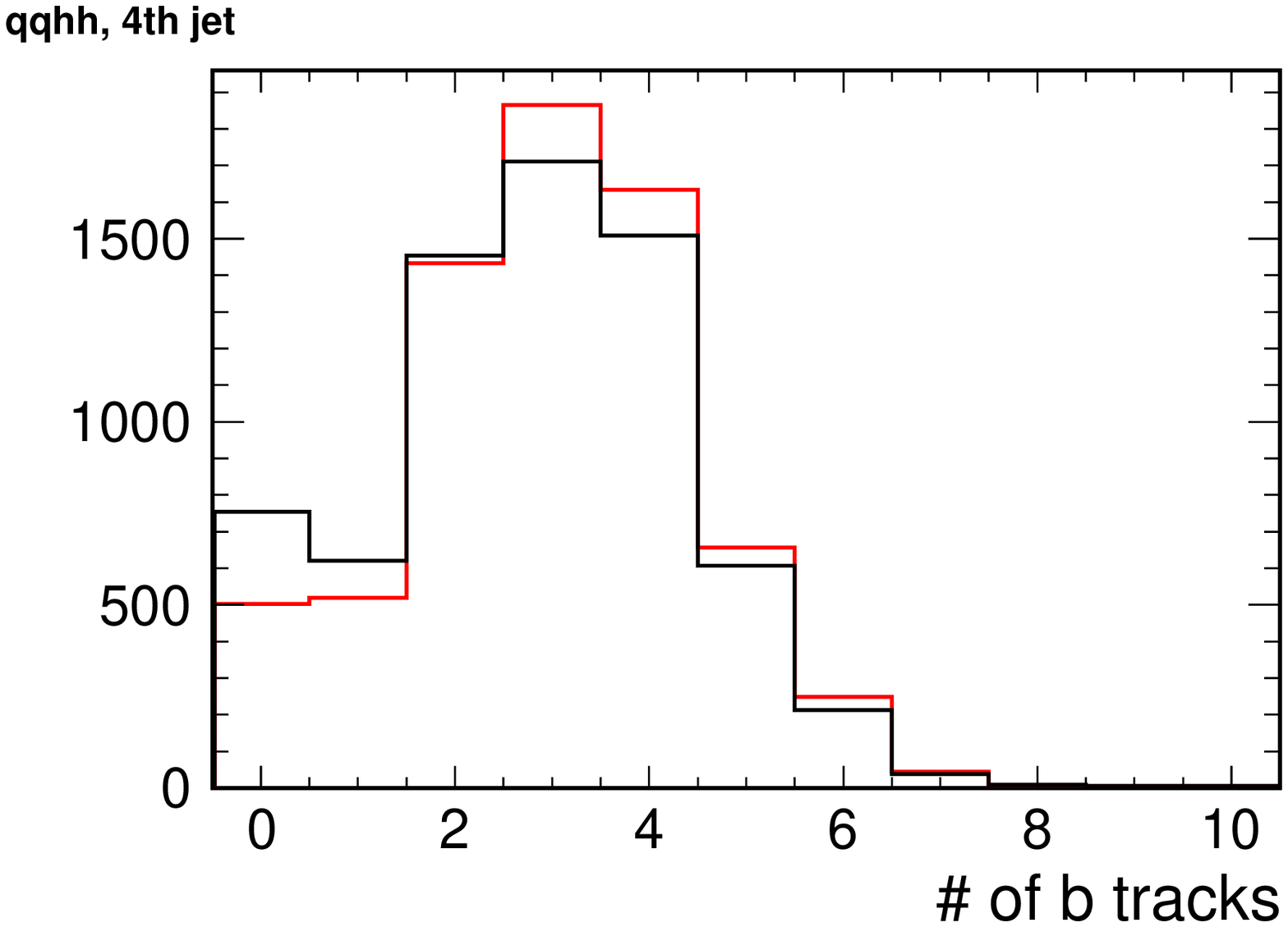}
        \end{center}
        \end{minipage}
\\
        \begin{minipage}[t]{.48\textwidth}
        \begin{center}
    \includegraphics[width=.8\textwidth]{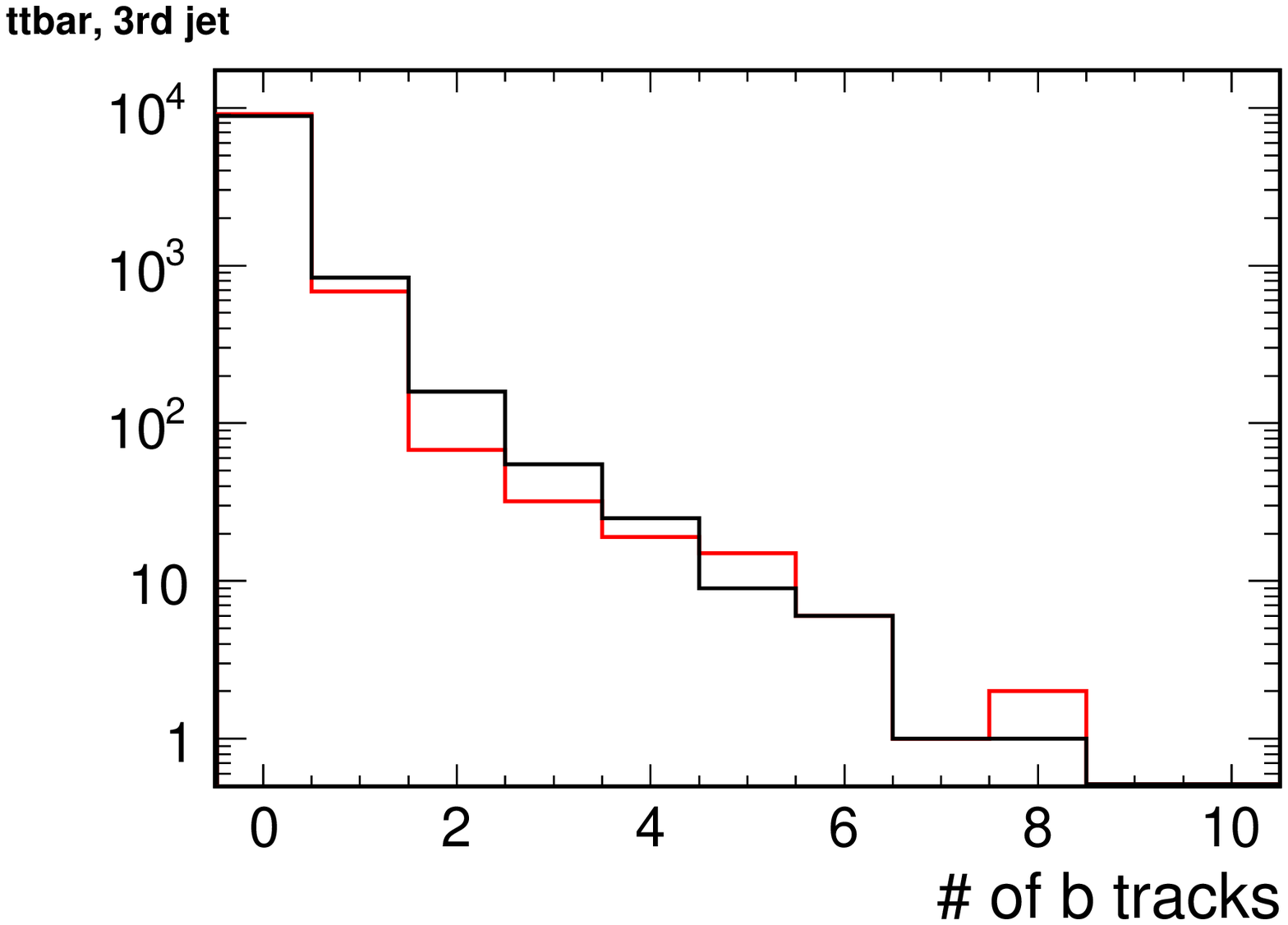}
        \end{center}
        \end{minipage}
        \begin{minipage}[t]{.48\textwidth}
        \begin{center}
    \includegraphics[width=.8\textwidth]{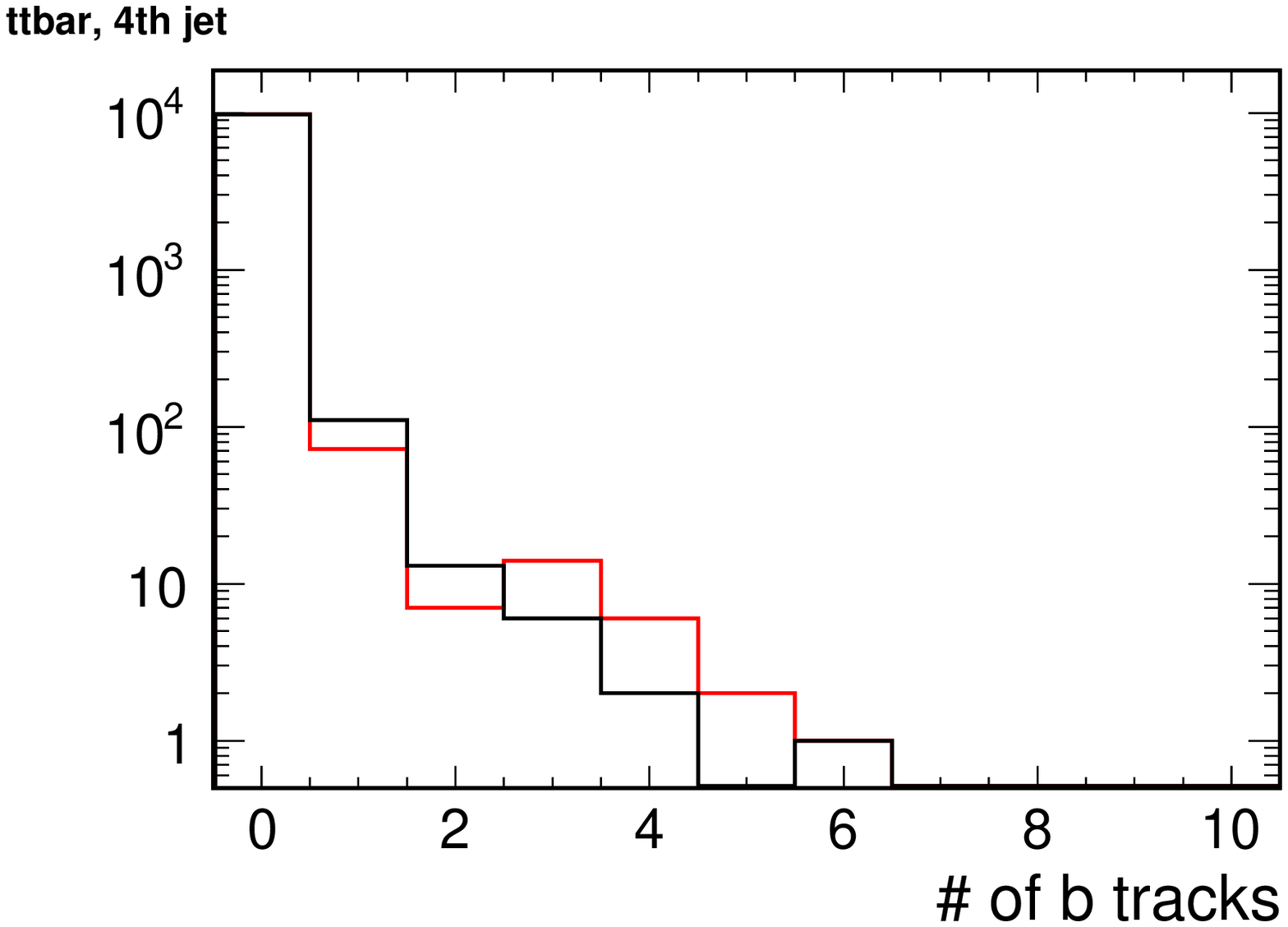}
        \end{center}
        \end{minipage}
        \end{center}
        \caption{Number of secondary tracks in the third (left) / fourth (right)
	  reconstructed jet with each method.
	  The upper two plots show the result for $ZHH \to qqbbbb$ events,
	  and the lower two plots show the result for $t\bar{t} \to bbcssc$ events.
	  The red lines show the result of our original algorithm and the black
	  lines show the result of Durham algorithm.}
        \label{fig:countbtracks}
\end{figure}

Figure~\ref{fig:countbtracks} shows the results.
The number of events with zero $b$ tracks in the third jet of the $bbcssc$ sample
is increased, which means better background rejection.
The number of events with non-zero $b$ tracks in the fourth jet of the $qqbbbb$ sample
is also increased,
which means better signal acceptance with our original method.

\subsection{$b$-tagging performance}

So far, we have used the MC generator information to compare the two algorithms.
Here, we show what an example of the difference in the reconstruction
through the performance of flavor tagging.
We use the LCFIVertex flavor tagging method~\cite{Bailey:2009ui}
which is applied after the jet clustering 
done by both the Durham algorithm and our original algorithm.
The output of LCFIVertex is the result of a artificial neural net 
which we will call $b$-likeness, given for each jet.
Since the $ZHH\to qqbbbb$ process
has 4 $b$ hadrons while $t\overline{t}\to bbcssc$ has 2 $b$ hadrons,
the $b$-likeness of the third and fourth jets,
in descending order of $b$-likeness,
is expected to be high for $qqbbbb$
and low for $bbcssc$.

\begin{figure}
        \begin{center}
        \begin{minipage}[t]{.3\textwidth}
        \begin{center}
    \includegraphics[width=.9\textwidth]{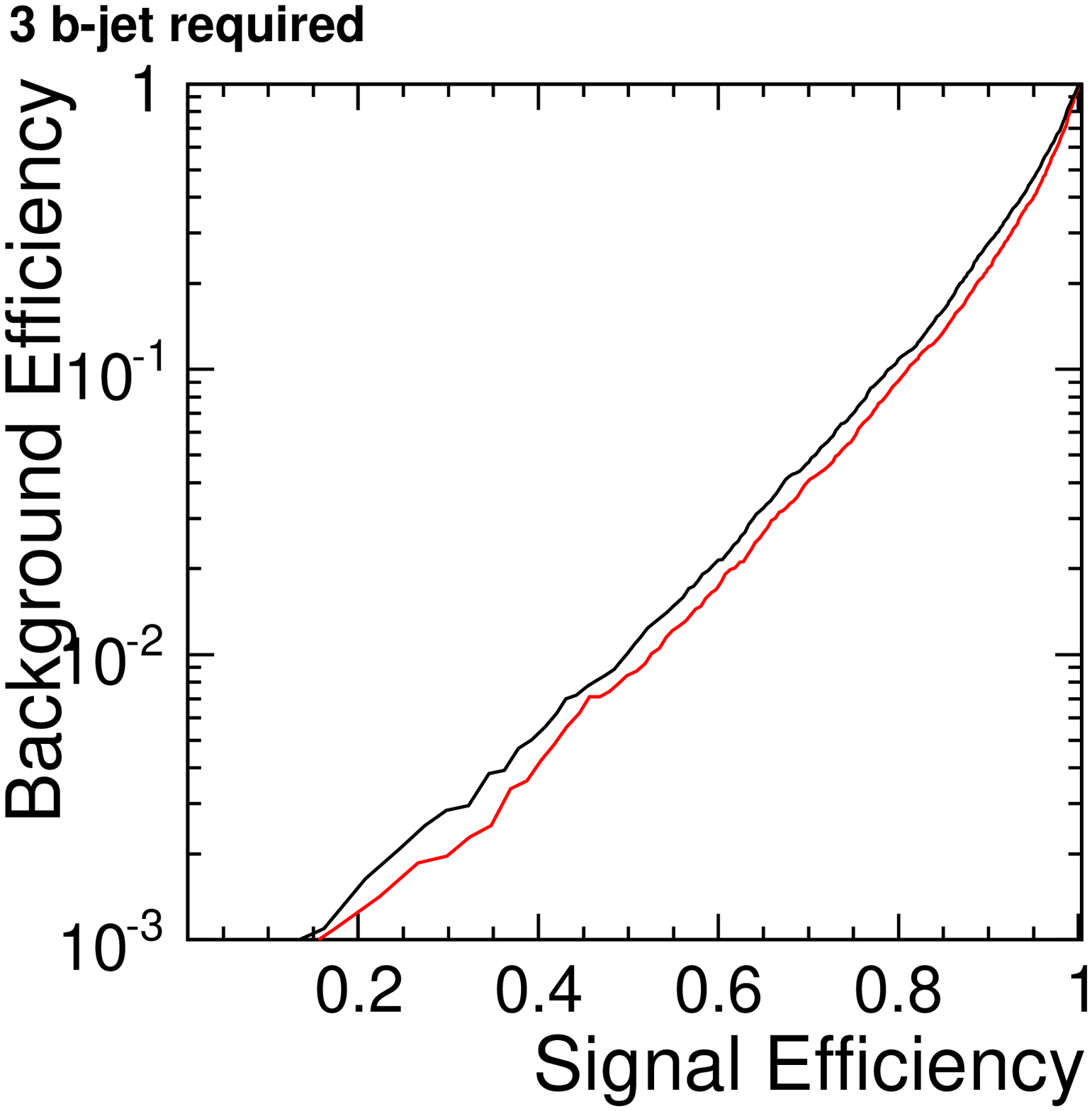}
        \end{center}
        \end{minipage}
        \begin{minipage}[t]{.3\textwidth}
        \begin{center}
    \includegraphics[width=.9\textwidth]{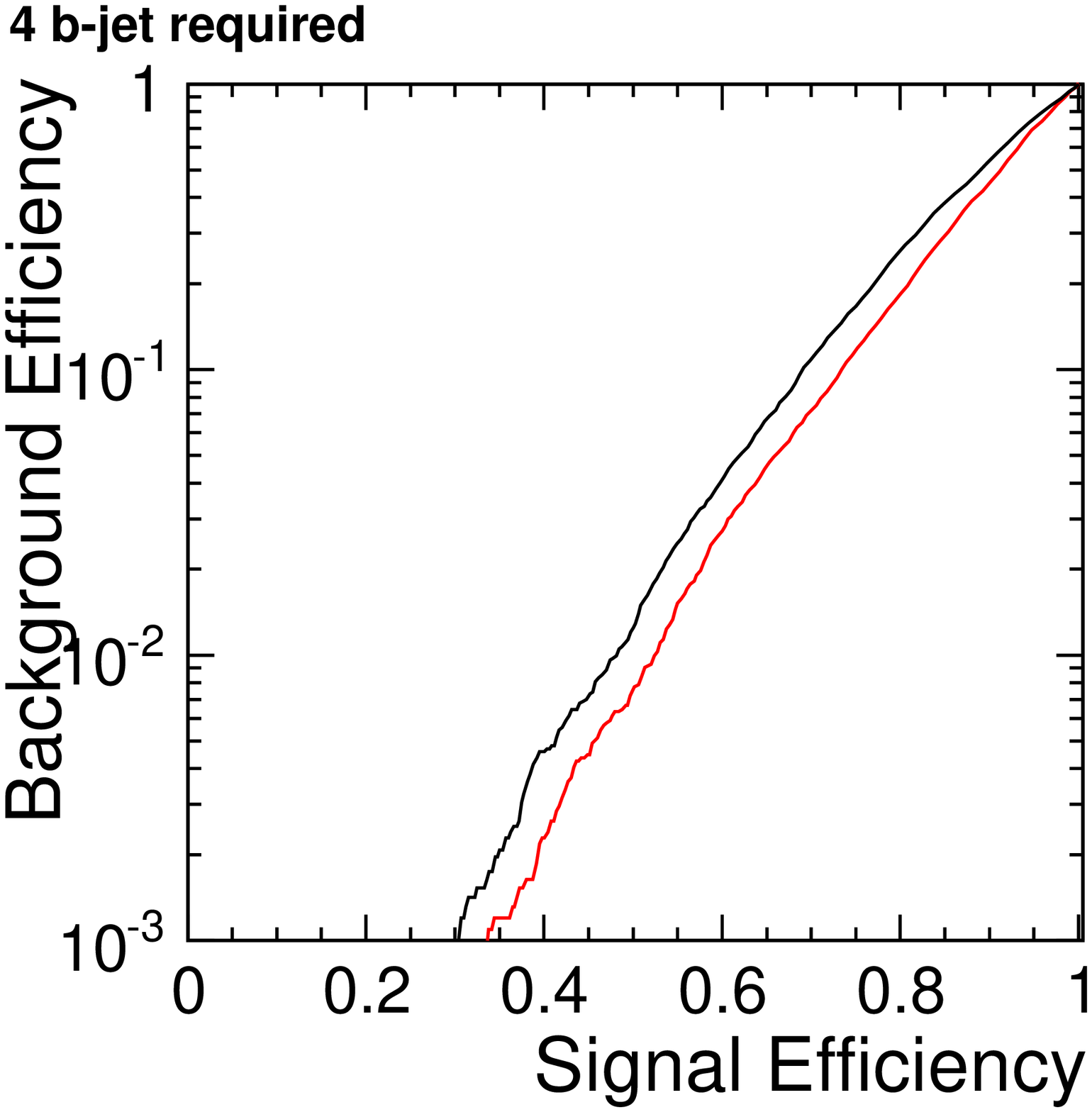}
        \end{center}
        \end{minipage}
        \begin{minipage}[t]{.3\textwidth}
        \begin{center}
    \includegraphics[width=.9\textwidth]{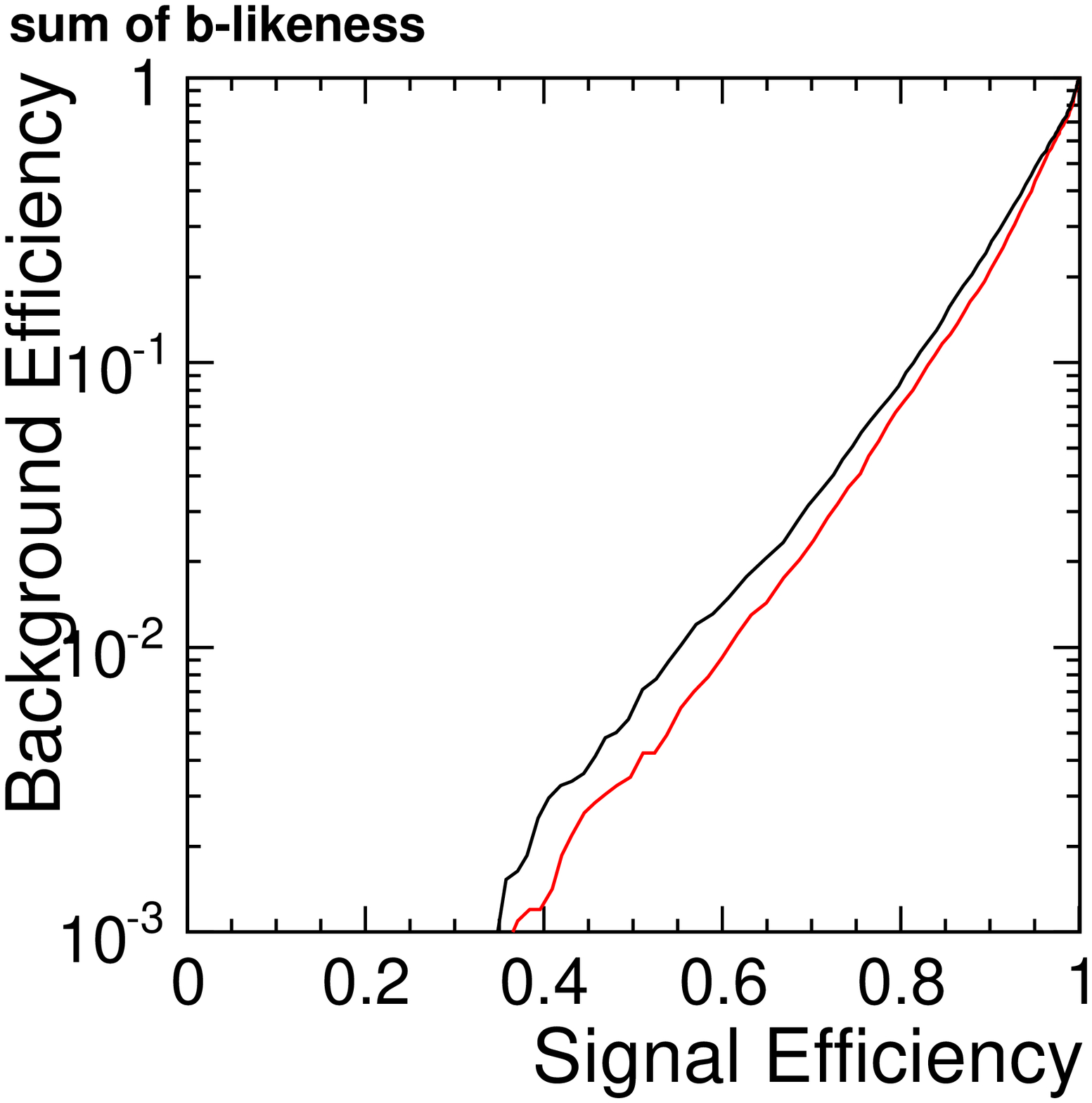}
        \end{center}
        \end{minipage}
        \caption{Comparison of the $b$-tagging performance for $ZHH \to qqbbbb$
	  and $t\bar{t} \to bbcssc$ events.
	  The horizontal axis of each plot shows the acceptance of $b$-tagging
	  in $ZHH \to qqbbbb$ events,
	  for the third jet (left), the fourth jet (center),
	  and the sum of the two (right),
	  with varying threshold of $b$-likeness.
	  The vertical axis shows the acceptance of $t\bar{t} \to bbcssc$ events
	  with the same threshold.
	  The red lines show the result with our original algorithm while the black line 
	  shows the result with the Durham algorithm.
}
        \label{fig:btag}
        \end{center}
\end{figure}

By changing the threshold of value of $b$-likeness
in defining the signal and background,
we obtain the efficiency plots shown in Figure \ref{fig:btag}.
In addition to the signal vs. background curve for the individual jets,
we also include the result of summing the two $b$-likeness
which combines the two information.
This result confirms that our method works at the reconstruction level as well.
It is worthwhile to note that the improvement over the Durham algorithm 
is significant in the high signal purity region,
with a background acceptance of less than 1\%.
Since the $ZHH$ analysis is known to need a powerful signal and background separation,
our algorithm is expected to significantly improve the sensitivity of the $ZHH$ analysis.

To illustrate the improvement in $b$-tagging, we perform a test event selection.
Here, we set the threshold of $b$-likeness such that the signal efficiency is
approximately 50\%.
We apply the selection individually to the third and the fourth jets,
as well as the combination of the two, and compare the differences.
Table~\ref{tbl:cut} shows the results.
With our original algorithm,
the number of remaining background events decreases by about 30\%
compared to the Durham algorithm.
Note that this number does not take into account the
correlations with other event selection criteria.
While we believe that our algorithm gives a significant boost to the
sensitivity of the $ZHH$ analysis,
a real demonstration must be performed in the context of the actual physics analysis.

\begin{table}
\begin{center}
\begin{tabular}{|l|r|r|r|r|r|r|r|}\hline
 & & \multicolumn{3}{c|}{Original} & \multicolumn{3}{c|}{Durham} \\ \cline{3-8}
 & No cut & 3rd jet & 4th jet & 3rd \& 4th jet & 3rd jet & 4th jet & 3rd \& 4th jet \\ \hline\hline
$ZHH \to qqbbbb$      & 8352 & 4233 & 4367 & 3163 & 4277 & 4382 & 3116 \\ \hline
$t\bar{t} \to bbcssc$ & 9930 &   95 &  113 &   20 &  145 &  137 &   29 \\ \hline
\end{tabular}
\caption{Comparison of the remaining number of events
  after the selection on the $b$-likeness for the $ZHH \to qqbbbb$
  and $t\bar{t} \to bbcssc$ events.
  The results with the individual selection on the third jet and the fourth jet
  are shown, as well as the combined selection.
  The selection threshold for the third and fourth jet is set
  such that the signal efficiency is approximately 50\%.}
\end{center}
\label{tbl:cut}
\end{table}

\section{Summary}

Jet clustering is an important tool
to discriminate physics processes involving many jets.
We have developed a new jet clustering algorithm
which employs the vertex information
in the context of a future linear collider such as the ILC.
The performance study targeted towards
an improve measurement of the Higgs self-coupling
in the six jet final states shows that the
separation between the $ZHH \to qqbbbb$ signal and
the $t\bar{t} \to bbcssc$ background
improves significantly
with our original jet clustering algorithm
when combined with the $b$-tagging information.
A more realistic performance check should be done
in the actual $ZHH$ physics analysis.





\bibliographystyle{elsarticle-num}
\bibliography{main}







\end{document}